\documentclass[journal=ancac3,manuscript=article,layout=onecolumn]{achemso}
\usepackage[version=3]{mhchem} 
\usepackage{siunitx}
\usepackage{graphicx}
\usepackage[label font=bf,labelformat=simple]{subfig}
\usepackage{caption}
\usepackage{chemformula}

\newcommand*\nc{NC}
\newcommand*\ncs{NCs}

\author{Marianna D'Amato}
\affiliation[LKB]{Laboratoire Kastler Brossel, Sorbonne Universit\'e, CNRS, ENS-PSL Research University, Coll\`ege de France, 4 place Jussieu, 75252 Paris Cedex 05, France}

\author{Lucien Belzane}
\affiliation[LKB]{Laboratoire Kastler Brossel, Sorbonne Universit\'e, CNRS, ENS-PSL Research University, Coll\`ege de France, 4 place Jussieu, 75252 Paris Cedex 05, France}

\author{Corentin Dabard}
\affiliation[INSP]{Sorbonne Université, CNRS - UMR 7588, Institut des NanoSciences de Paris, INSP, F-75005 Paris, France}

\author{Mathieu Silly}
\affiliation[SOLEIL]{Synchrotron SOLEIL, L'Orme des Merisiers, Départementale 128, 91190 Saint-Aubin, France}

\author{Gilles Patriarche}
\affiliation[C2N]{Centre de Nanosciences et de Nanotechnologies, CNRS, Université Paris-Saclay, C2N, Palaiseau 2110, France}

\author{Quentin Glorieux}
\affiliation{Laboratoire Kastler Brossel, Sorbonne Universit\'e, CNRS, ENS-PSL Research University, Coll\`ege de France, 4 place Jussieu, 75252 Paris Cedex 05, France}

\author{Hanna Le Jeannic}
\affiliation{Laboratoire Kastler Brossel, Sorbonne Universit\'e, CNRS, ENS-PSL Research University, Coll\`ege de France, 4 place Jussieu, 75252 Paris Cedex 05, France}

\author{Emmanuel Lhuillier}
\affiliation[INSP]{Sorbonne Université, CNRS - UMR 7588, Institut des NanoSciences de Paris, INSP, F-75005 Paris, France}

\author{Alberto Bramati}
\affiliation{Laboratoire Kastler Brossel, Sorbonne Universit\'e, CNRS, ENS-PSL Research University, Coll\`ege de France, 4 place Jussieu, 75252 Paris Cedex 05, France}
\email{alberto.bramati@lkb.upmc.fr}

\title[]
  {Highly photostable Zn-treated halide perovskite nanocrystals for efficient single photon generation}


\begin{document}
\begin{abstract}
Achieving pure single-photon emission is essential for a range of quantum technologies, from optical quantum computing to quantum key distribution to quantum metrology. Among solid-state quantum emitters, colloidal lead halide perovskite (LHP) nanocrystals (NCs) have garnered significant attention due to their interesting structural and optical properties, which make them appealing single-photon sources (SPSs). 
However, their practical utilization for quantum technology applications has been hampered by environment-induced instabilities. In this study, we fabricate and characterize in a systematic manner
Zn-treated \ch{CsPbBr_3} colloidal NCs obtained through \ch{Zn^{2+}} ion doping at the Pb-site, demonstrating improved stability under dilution and illumination.
These doped NCs exhibit high single-photon purity, reduced blinking on a sub-millisecond timescale and stability of the bright state for excitation powers well above the saturation levels. Our findings highlight the potential of this synthesis approach to optimize the performance of LHP-based SPSs, opening up interesting prospects for their integration into nanophotonic systems for quantum technology applications. 
\end{abstract}

\section{Introduction}
 
Single photon sources (SPSs) play a fundamental role in numerous quantum technology applications, including quantum computing\cite{o2009photonic,northup2014quantum,kok2007linear}, quantum simulation\cite{georgescu2014quantum,harris2017quantum}, quantum metrology\cite{dowling2008quantum,giovannetti2006quantum} and quantum communication\cite{lo2014secure,wang2015quantum}.
Considered ideal flying qubits, they offer easy manipulation and detection as well as long-range transmission. Ideal SPSs must exhibit high efficiency in single photon emission, as well as high single photon purity and indistinguishability. Ideally, a SPS should  exhibit a narrow emission linewidth together with fast emission lifetimes to achieve high data rates for quantum information processing\cite{buckley2012engineered}.   
while operating at room temperature \cite{zhou2018room}, for practical usability.
Viable candidates for SPSs include solid-state emitters, such as organic molecules\cite{toninelli2021single,lounis2000single}, color centers in diamonds\cite{kurtsiefer2000stable,neu2011single}, and epitaxially-grown and colloidal quantum dots (QDs)\cite{somaschi2016near,michler2000quantum}. Solid-state SPSs offer distinct advantages, such as the ability to generate single photons on-demand, unlike heralded single photon sources based on parametric conversion processes, while being brighter than atoms and ions. Furthermore, solid-state SPSs demonstrate the potential for integration into nanophotonics architectures.
Amongst various quantum dots, colloidal lead halide perovskite (LHP) nanocrystals (NCs) of the \ch{APbX_3} form (with A = \ch{CH_3NH_3}, Cs, or FA and X = Cl, Br, or I) have recently garnered significant attention in the quantum optics community due to their remarkable structural and optical properties which make them appealing as single photon sources.\cite{park2015room,raino2016single,huo2020optical,utzat2019coherent,soci2023roadmap}.
LHP NCs possess peculiar electronic structure \cite{canneson2017negatively,ramade2018fine,fu2017neutral} compared with the well known II-VI and even III-V QDs, although the exact nature of their dark state is still a subject of debate.\cite{sercel2019exciton,tamarat2020dark,tamarat2019ground,ramade2018fine,becker2018bright}.
This distinctive feature contributes to their defect-tolerant nature\cite{huang2017lead}, enabling them to achieve remarkably high photoluminescence (PL) quantum yields (QY), reaching values as high as $95-100\%$ from shell-free nanoparticles\cite{akkerman2018genesis}. Additionally, they offer tunable absorption and emission spectra across the visible range by adjusting their size and composition, narrow emission linewidths
 (12-50 nm), and relatively short PL lifetimes 
 (1-29 ns)\cite{protesescu2015nanocrystals, d2023color} at room temperature. Notably, their ease of fabrication using well-established wet-chemistry techniques adds to their appeal, making them a cost-effective option.
However, due to their ionic lattice, LHP are very sensitive to water and moisture and, in addition, they are subject to photon-induced structural reorganization \cite{brivio2016thermodynamic,siegler2022water,ouyang2019photo,li2018ultraviolet}. These drawbacks can severely limit their use for quantum technology applications which require a high degree of stability which is highly linked to the indistinguishability and coherence of the single photons.
Large efforts have therefore been dedicated to increase the LHP stability either through better ligand passivation\cite{zhang2020stable,krieg2018colloidal,krieg2019stable,swarnkar2015colloidal}, by growing a surrounding shell\cite{an2021low,palei2020robustness,loiudice2019universal} or by embedding the NCs in a protective matrix to prevent exposure to moisture\cite{raja2016encapsulation}. 
Recently, the doping of the Pb-site cation in the perovskite lattice with metal ions (\ch{Sn^{2+}}, \ch{Cd^{2+}}, \ch{Zn^{2+}}) has been considered as one of the most effective methods to improve the structural stability and the optical performance of LHP-based light-emitting diodes (LEDs) and solar cells\cite{ding2019stable,bi2020stable}.
In perovskite \ncs{} glasses for white light-emitting diodes (WLEDs) applications, a small amount of ions doping improved the Goldschmidt tolerance factor, i.e. an empirical index to predict the perovskite structural stability\cite{li2016stabilizing}, without changing the perovskite crystalline form\cite{ding2019stable,thapa2019zn}. 
In particular, \ch{Zn^{2+}} is of significant interest due to its non-toxicity, its high stability against oxidation or reduction compared to other dopants, and its effectiveness in eliminating defect states - contributing to passivate halide vacancy defects on the surface and reduce grain boundary surfaces\cite{kooijman2019perovskite} - where non-radiative recombination typically occurs\cite{pols2022happens}. As a result, \ch{Zn^{2+}} has a beneficial impact on both the efficiency and long-term stability of the system\cite{shen2019zn}.

Here we focus on the \ch{Zn^{2+}} doping of \ch{CsPbBr_3} \ncs{} to achieve Zn-treated \ch{CsPbBr_3} \ncs{} with improved stability and brightness compared to the pristine particles while preserving their excellent quantum properties.

 \section{Results and discussion}

\begin{figure*}[ht!]
     \includegraphics[width=140 mm]{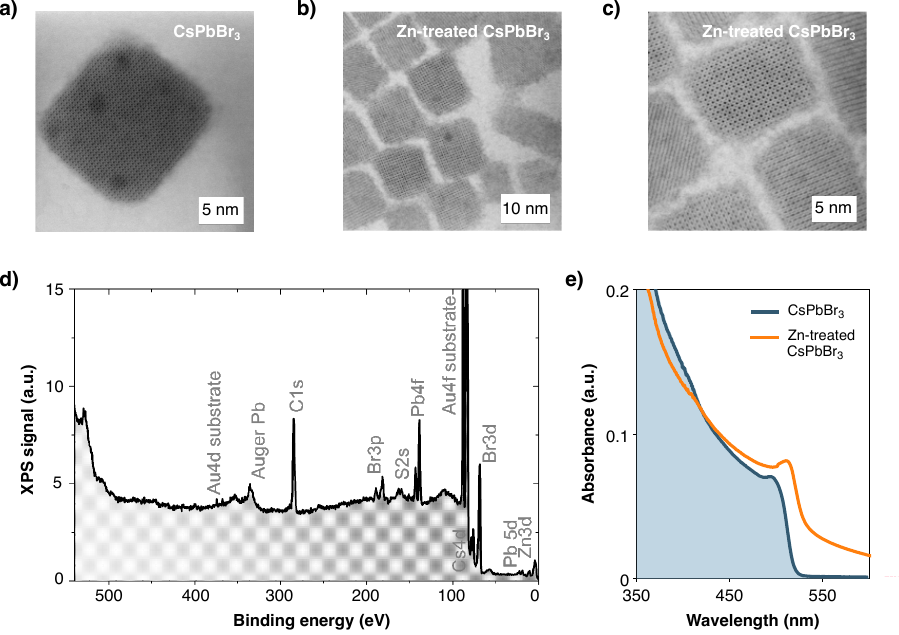}
     \caption{ Zn-treated \ch{CsPbBr_3} nanocrystals. 
      a) High magnification HRTEM image of the pristine \ch{CsPbBr_3} \ncs. Low (b) and high (c) magnification HRTEM images of the Zn-trated \ch{CsPbBr_3} \ncs. d) X-ray photoemission survey spectrum for the Zn-trated \ch{CsPbBr_3} \ncs. e) Absorption spectra for the pristine \ch{CsPbBr_3} \ncs{} before and after their exposure to Zn(DDTC) precusor.}
   \label{fig:lhu}
\end{figure*}

 To synthesize the Zn-treated \ch{CsPbBr_3} \ncs, we start by growing pristine \ch{CsPbBr_3} \ncs{} using the procedure developed by Protecescu et al. \cite{protesescu2015nanocrystals}, obtaining cubes with 12 nm edge, as shown in  high resolution transmission electron microscopy (HRTEM) images in Figure\ref{fig:lhu}a. Subsequently, the grown cubes are mixed with a zinc and sulfur containing molecular precursor as proposed by Ravi et al.\cite{ravi2020cspbbr3} (see Methods in Supporting Information for full details regarding the growth).  However, the procedure also comes with a significant increase in particle size which ends up loosing quantum confinement. Such large particle with size around \SI{50}{nm} get less likely to behave as single photon emitter. This is why we perfom a size selection step to discard all the largest particle and preserving the smallest. The morphology of the resulting Zn-treated \ch{CsPbBr_3} \ncs{} is investigated using HRTEM, as shown in Figure\ref{fig:lhu}b-c, confirming that the preserved NCs maintain their original pristine size, without any noticeable evidence of shell growth, as opposed to what observed by Ravi et al.\cite{ravi2020cspbbr3} on large nanoparticles. By performing  X-ray photoemission spectroscopy (Figure\ref{fig:lhu}d and 
 FigureS2) and energy dispersive X-ray spectroscopy (FigureS3),  we confirm that a low amount  ($0.5\%$) of Zn has been incorporated in the NCs,  which is consistent with previous report relative to Zn doping and alloying of perovskite \ncs{}\cite{swarnkar2018can,bi2019thermally,mondal2018achieving,zhou2019ag}. Figure\ref{fig:lhu}e shows the absorption spectra of the pristine and Zn-treated \ncs, displaying respectively an absorption edge at around 520 nm and 530 nm.

\subsubsection{Photostability under dilution and illumination}

The colloidal stability of perovskite \ncs{} 
mainly depends on their surface chemistry, due to their high surface-to-volume ratio. The highly dynamic binding between capping ligands\cite{fiuza2023ligand} and the ionic NC lattice can cause the detachment of ligands and halide atoms from the surface, leading to a disordered surface with defects that can reduce the long-term stability. Indeed, these defects can act as trap sites for optically excited charge carriers and excitons, leading to non-radiative recombination processes and a significant reduction in the photoluminescence quantum yield (PLQY). Furthermore, certain defect traps may undergo photochemical reactions, such as 
photo-oxidation, which can cause the degradation and structural collapse of the NC, 
and this is further accelerated in the presence or moisture\cite{siegler2022water}.
For applications in photonic devices and integrated single-photon sources, high colloidal stability of the NCs is essential to isolate a single emitter through high dilution\cite{pierini2020highly}, and this requires overcoming environmental-induced instabilities.

\begin{figure}[ht!!]
    \centering
    \includegraphics[width=80 mm]{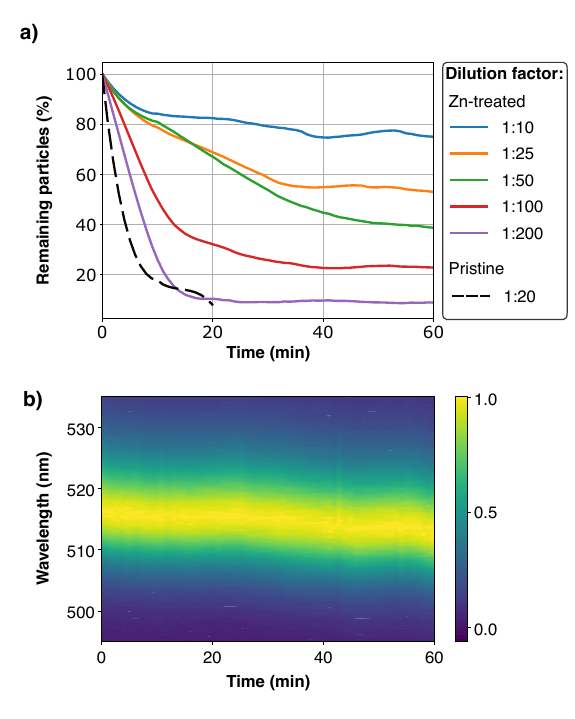}
     \caption{ Photostability of Zn-treated \ch{CsPbBr_3} \ncs{}. 
       a)~ In colors percentage of Zn-trated \ch{CsPbBr_3} \ncs{} still emitting after 1 hour under strong LED illumination as a function of  dilution. 1:x, with x going from 10 to 200, meaning x times dilution factor from the original solution. In dashed black line, percentage of pristine \ch{CsPbBr_3} \ncs{} still emitting after 20 minutes in a 1:20 dilution. 
       b)~ Evolution of the CEW from a single Zn-treated \ch{CsPbBr_3} \nc{} excited at its saturation intensity. Each spectrum is normalized.}
   \label{fig:photostability}
\end{figure}

We assessed the stability of NCs under dilution by studying NCs ensembles using wide-field microscopy\cite{krieg2019stable}. To this end, we prepared five samples by diluting a NCs solution  with a molar concentration of approximately \SI{1}{\micro M} with toluene at various dilution ratios, ranging from $1\colon10$ to $1\colon200$, and spin-coating a droplet of \SI{30}{\micro l} of each dilution on a glass coverslip. In order to investigate the evolution of photobleaching with dilution, the samples were exposed to intense light from a LED (\SI{400}{nm})  in a wide-field configuration (see the setup description in Section SIII of Supporting Information). The sample  emission was recorded, with a frame taken every \SI{20}{s} over a period of \SI{1}{h}. 
By counting the number of emitting NCs in each
frame, we can track the temporal evolution of ensemble emission. Zn-trated \ch{CsPbBr_3} NCs demonstrate exhibit superior resistance to dilution compared to pristine \ch{CsPbBr_3}. For the latter ones photobleaching occurs within \SI{5} minutes regardless of the dilution used, as previously reported\cite{pierini2020highly}. This behaviour is reported in Figure\ref{fig:photostability}a by the dashed dark line, representing a 1:20 dilution and serving as a basis for direct comparison. In the case of Zn-trated NCs, represented by colored lines in the figure, a dilution ratio of 1:10 exhibits remarkable stability, with over $70\%$ of particles still emitting after 1 hour. For the $1\colon20$ dilution, approximately $55\%$ of particles continue to emit after 1 hour, followed by $40\%$ for a $1:50$ dilution and $20\%$ for a $1\colon100$ dilution. Only when reaching a $1:200$ dilution do we start to observe the bleaching effect, with a bleached time of around \SI{15} {min}. In this case, a constant residual photoluminescence is observed, which can be attributed to the presence of small aggregates. The improved resistance to dilution in Zn-treated \ch{CsPbBr_3} perovskites may be attributed to a reduction in defect states resulting from the doping process. This reduction can contribute to the diminished degradation of the NCs, even without polymer encapsulation\cite{raino2019underestimated} and in contact with air.

Moving to a confocal microscopy scheme, we monitored then the spectral stability of individual NCs in the $1:10$ dilution, which allows us to address an individual emitter, as 
confirmed by single photon purity 
(see next section),
and collect its luminescence A single NC was excited with a pulsed laser at 405 nm (pulse width $<100$ ps and repetition rate of 2.5 MHz), as described in the Section SIII of Supporting Information, for \SI{1}{h} while its emission spectrum was collected every $5$ minutes.
When conducting measurements under ambient conditions, a continuous blue-shifting of the photoluminescence (PL) spectrum has been frequently reported in the literature\cite{raja2016encapsulation,huang2017lead,pierini2020highly}. This is a typical signature of photo-induced degradation that leads to a layer-by-layer etching of the NC surface, resulting in a reduction in the size of the QDs and an increase in the band-gap energy\cite{an2018photostability}. 
Figure\ref{fig:photostability}b shows the 2D colored plot of consecutive PL spectra obtained, showing a blue-shift of the central emission wavelength (CEW) of less than 4 nm in \SI{1}{h}. 
The photostability of these emitters exhibits therefore significant improvements compared to previously reported results in the literature for emitters without encapsulation where, typically, a CEW's blue-shift of more than 10 nm is observed within a few tens of seconds\cite{raino2019underestimated}. 

\subsubsection{Single photon purity}

\begin{figure*}[ht!]

    \includegraphics[width=140 mm]{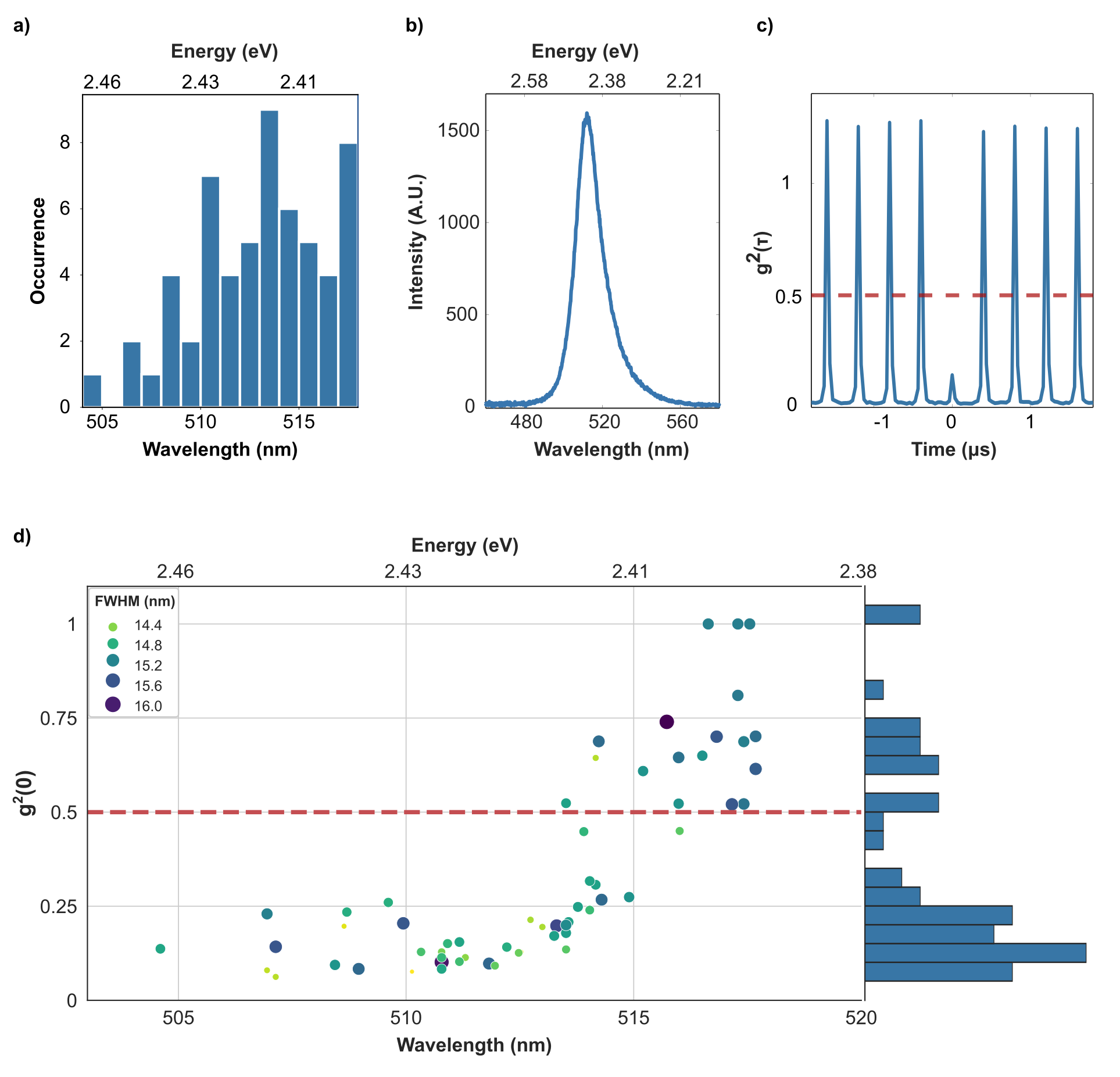}
     \caption{ Single photon purity.  
      (a) Histogram of the measured CEWs for a set of 64 \ncs{}. (b) Representative emission spectrum of a single NC. (c) Example of second-order correlation function of a single NC emitting high-quality single photons ($g^2(0)=0.08$ after background counts subtraction). The red-line indicates the threshold for single-photon emission claim.
      (d) Measured $g^2(0)$ values as a function of the CEW for the 64 \ncs{}. All the emitters were excited at their saturation intensity. }
   \label{fig:singlephotonpurity}
\end{figure*}

In this work, we thoroughly analyzed the optical and quantum properties of a set of 64 Zn-treated 
\ch{CsPbBr_3} emitters. To establish a reference for the characterization of multiple emitters, all measurements were performed at the specific saturation power of each nanocrystal. Figure\ref{fig:singlephotonpurity}a presents the CEWs distribution of the 64 emitters. Due to the inhomogeneus slight size distribition of the NCs, the CEW ranges from 504 to 518 nm (2.46 to 2.40 eV), with an average value of \ch{512.5  ±  3.3} nm.

A typical emission spectrum is reported in
Figure\ref{fig:singlephotonpurity}b , with a CEW of $512$ nm and a full width half maximum (FWHM) of $15$ nm. 
To verify that the Zn-treated \ch{CsPbBr_3} NCs behave as sources of quantum light, we evaluate their single photon purity measuring the second order correlation function\cite{brown1956correlation}:
$$
g^{2}\left( \tau \right)= \dfrac{\left<I(t)I(t+\tau)\right>}{\left< I(t) \right>^2},
$$
where $I$ is the intensity of the emission, $t$ the time and $\tau$ the delay time between two photon detection events. The value of $g^2(0)$ gives the probability of two photons being emitted simultaneously by the source. Figure\ref{fig:singlephotonpurity}c presents a typical histogram of $g^2(\tau)$, showing a photon antibunching value of $g^2(0)=0.08$ after background subtraction, as explained in Section SIV of Supporting Information.
We measured the $g^2(\tau$) for all 64 emitters and analyzed the $g^2(0)$  evolution as a function of emission wavelength. As the emission wavelength increases with the nanoemitter size, this measurement allows to exploring the effect of quantum confinement on single photon purity. 
 The results are displayed in  Figure\ref{fig:singlephotonpurity}d. We observe that the $g^2(0)$ increases from $0.1$ to $1$ as the emission wavelength increases, consistently with the trend reported for \ch{CsPbBr_3} perovskite
 NCs\cite{pierini2020highly,zhu2022room}.
The observed trend strongly indicates that an increase of the degree of quantum confinement corresponds to an increase of the single-photon purity. Below approximately 515 nm, around $94\%$ of the NCs exhibit antibunching behaviour. However, beyond this threshold, only a single nanocrystal was observed to emit single photons, while broader FWHM values become more common. These findings suggest that for emission wavelengths exceeding approximately 515 nm, the quantum confinement regime is no longer maintained.

\subsubsection{Blinking analysis}

\begin{figure*}[ht!]

    \includegraphics[width=\linewidth]{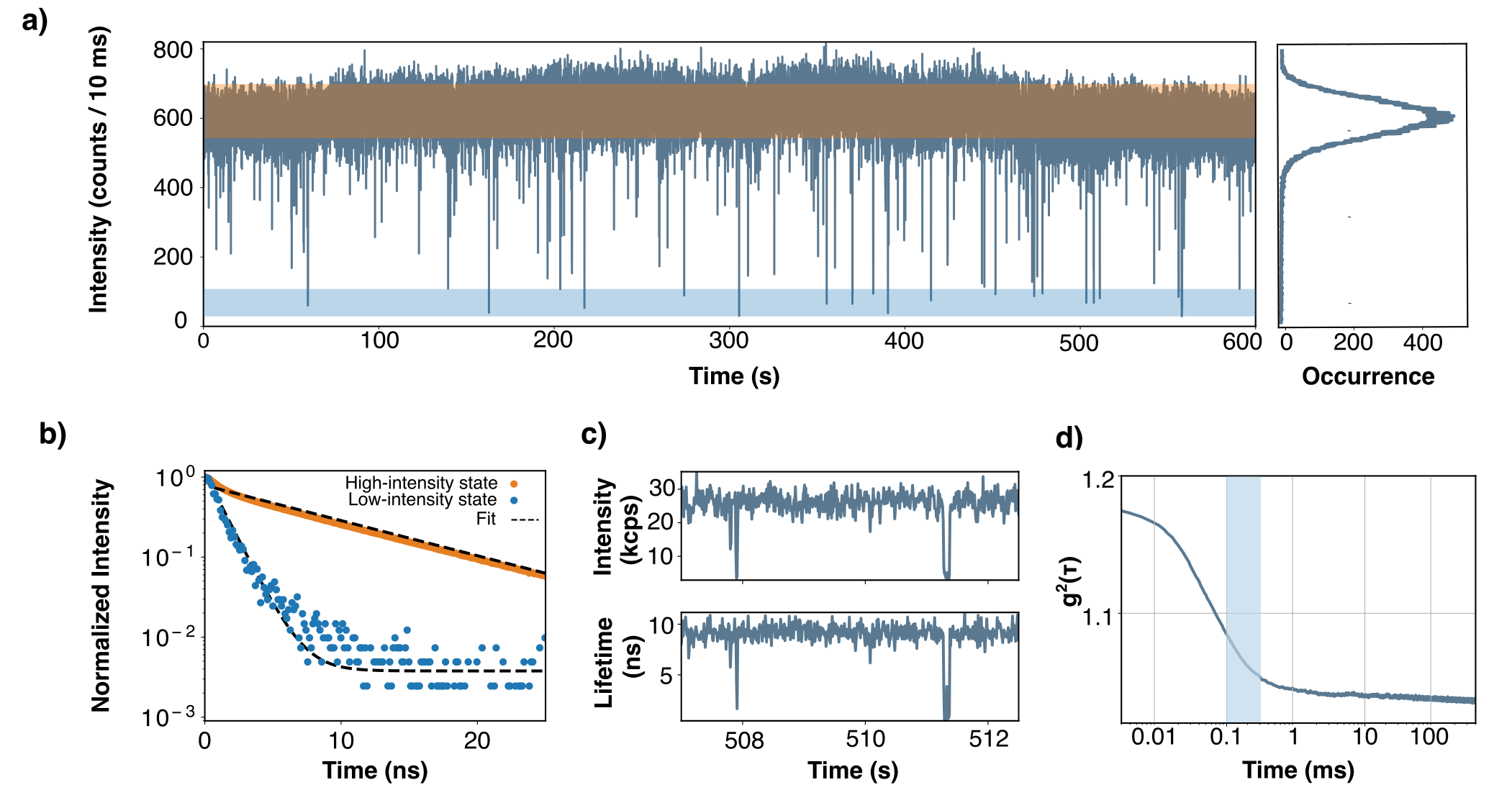}
     \caption{ Blinking dynamics.  
      (a)Intensity time trace of a single \nc{}, and corresponding relative histogram of intensity occurences.  
      (b) Decays for the low-intensity states (blue) and high-intensity states (orange). (c) Zoom-view of the intensity time-trace (upper box) and lifetime (lower box) of a single \nc{}. (d) $g^2(\tau)$ at large time delays for a single \nc{}.}
   \label{fig:blinking}
\end{figure*}

Figure\ref{fig:blinking}a shows a typical PL intensity time-trace for a single Zn-treated \ch{CsPbBr_3} NC, together with the corresponding intensity histogram, obtained through a time-tagged time-resolved (TTTR) method with a temporal resolution of 126 ps (see Section SIII in Supporting Information for details on the method). The measurement was carried out at the saturation power with a binning time of 10 ms. The results show that the NC maintains a highly stable and bright emissive state, as evidenced by a constant count rate over the entire \SI{600}{s} integration time. Moreover, as compared to the pristine NCs (see FigureS6 in Supporting Information), the brightness of the Zn-treated \ch{CsPbBr_3} NCs is improved. To investigate the nature of the blinking, two intensity windows were selected from the intensity time-trace, corresponding to the high-intensity and low-intensity states (see orange and blue areas, respectively, in Figure\ref{fig:blinking}a). The corresponding TTTR signals were then pinpointed to retrieve their photoluminescence (PL) decays, which where compiled for statistics. The results are shown in Figure~\ref{fig:blinking}b. The PL decays for the two intensity windows were fitted with a mono-exponential model, taking into account the background noise, yielding lifetimes of $\tau_1=10.2$ ns and $\tau_2=1.3$ ns, respectively for the high-intensity and low-intensity states. The overall PL decay was fitted using a bi-exponential model that accounted for background noise, 
as depicted in FigureS5 in Supporting Information, along with the corresponding $g^{2}$ curve confirming high single-photon purity.

According to the charging/discharging model\cite{galland2011two}, the high-intensity states can be attributed to radiative excitonic recombinations, while the faster decay of the low-intensity states is indicative of emission from charged excitons (trions). Trions have higher likelihood to experience non-radiative Auger recombination\cite{becker2018bright} resulting in a reduced PL.

Figure\ref{fig:blinking}c depicts a zoomed-in view of a 7-seconds segment of the intensity time-trace, with a bin time of 10 ms. In the lower panel of Figure\ref{fig:blinking}c, the average lifetime of photons within each bin is plotted as a function of time. Notably, a clear correlation between the photon lifetime and emission intensity is observed, which is indicative of type A blinking behavior. In this type of blinking, the lifetime of the emitted photons is expected to depend on the emission intensity, in accordance with the charging/discharging model\cite{galland2011two}.

 Blinking dynamics are commonly investigated on time scales of milliseconds and longer by means of binning and thresholding\cite{kuno2001off} or change-point analysis\cite{bae2016understanding}. 
 Nevertheless, these approaches rely on the binning of photon detection events and may not provide reliable information, especially in the case of fast blinking. In this work, we employed the second-order correlation function at large time delays, which allows to accurately assess the amplitude and rate of blinking at short time scales that would not be accessible through binning the signal\cite{manceau2014effect,manceau2018cdse}. The $g^2(\tau)$ function is shown over a wide range of time scales, from 10 ns to around \SI{1}{s}, as reported in Figure\ref{fig:blinking}d. Notably, at short delays ($\tau<$ \SI{10}{\micro s}) the $g^2(\tau)$ function exhibits a super-Poissonian bunching value of $1.18$, due to the flickering between the two intensity states\cite{messin2001bunching}. 
 For delays above \SI{100}{\micro s}, the $g^2(0)$ value decreases towards unity, meaning that switching between the two states does not happen on these longer time-scales. In comparison to pristine NCs, showing an higher $g^2(0)$ value at short delays and a millisecond bliking time-scale as shown in Figure S6 of Supplementary Information, Zn-treated NCs exhibit significantly diminished $g^2(\tau)$ blinking-induced bunching amplitude, corresponding to a strongly reduced blinking probability, and to faster blinking rates on a microsecond time scale.

\subsubsection{Stability of the single photon emission over the power}

For an ideal two-level system, emitting one photon per excitation pulse, the emitted intensity would show a perfect saturation as a function of the excitation power and the $g^2(\tau)$ function a perfect antibunching $g^2(0)=0$ independent of the excitation power. On the other hand, in real solid state emitters, in which multi-excitonic states can relax radiatively, the saturation curve is non perfect, the emitted light can exhibit a linear trend with the excitation power and the antibunching is degradeted at high excitation power.
In confined LHPs NCs under high excitation power,  the contribution of multiexciton states to the emission, although  significantly reduced  by an efficient Auger non-radiative recombination\cite{zhu2022room} results in a non-perfect saturation curve and a power dependent $g^2(0)$ value.

\begin{figure*}[ht!]
    \includegraphics[width=\linewidth]{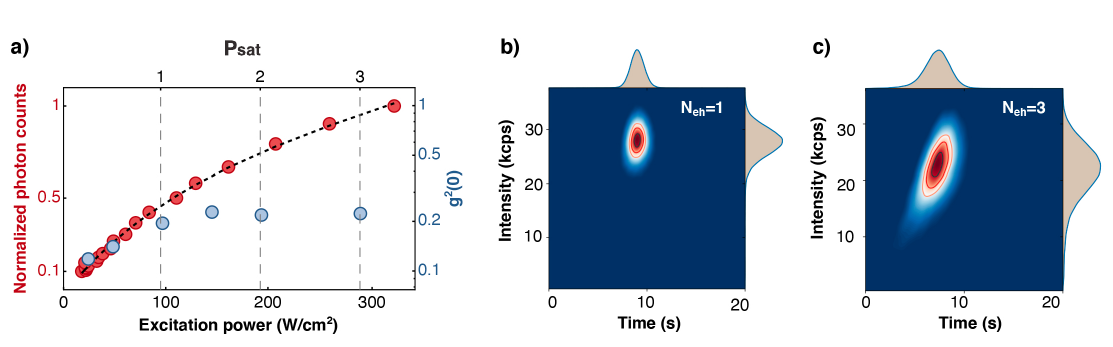}
     \caption{Stability of the single photon emission versus the excitation power. (a) In red: saturation measurement of a single
\nc{}. The dots are the experimentally measured counts with the excitation power. The dashed black line is the fitting function from eq\ref{eq:saturation} . In blue: evolution of the antibunching $g^2(0)$ values measured for different excitation powers (respectively to $0.25,0.5,1,1.5,2,3$ times the $P_{sat}$).(b,c) Fluorescence lifetime-intensity distribution (FLID) images of a single emitter, excited at \ch{P_{sat}} (c) and
\ch{3 P_{sat}} (d).}
   \label{fig:Figure5}
\end{figure*}

Figure\ref{fig:Figure5}a shows the PL intensity of a single Zn-treated \ch{CsPbBr_3} \nc{} measured as a function of the excitation power.
The data were fitted with the following model:
\begin{equation}
\label{eq:saturation}
    I=A\cdot \bigg[ 1- e^{-\frac{P}{P_{sat}}}\bigg] + B \cdot  P
\end{equation}
where \ch{P_{sat}} is the saturation power (i.e. the excitation power for which the number of excitons \ch{N_{eh}= 1}), and A and B are 
two constants that depend respectively on the intensity of the single- and bi-exciton components of the emission. The saturation power was extracted from the fitting curve and used for single particle measurements. We measured the $g^2(\tau)$ at different excitation power 
(respectively to $0.25,0.5,1,1.5,2,3$ times the 
\ch{P_{sat}}), observing that the $g^2(0)$ is increasing below \ch{P_{sat}} and remains constant for higher excitation powers. This indicates that, for the studied \ncs{}, multi-excitonic events are negligible. To evaluate the stability of the bright state, we use a Fluorescence Lifetime-Intensity Distribution (FLID) analysis, which provides a visual way to analyze correlations between photoluminescence intensities and lifetimes\cite{galland2011two}. 
The FLID distributions in Figure\ref{fig:Figure5}b,c display the occupation, in a two-dimensional lifetime-intensity space of a specific state, for the excitation powers respectively of \ch{P_{sat}} and \ch{3 P_{sat}}. In particular, for \ch{P_{sat}} we observe that the emission remains in the bright state characterized by significantly reduced blinking and the absence of photobleaching. For 
\ch{3 P_{sat}}, we begin to observe an elongated and spread shape as result of a slight increase of photobleaching and blinking processes (see Figure S7 of the Supporting Information for the complete shape evolution). Comparatively, pristine \ncs{} already showed an important spread at \ch{P_{sat}}(see Figure S6). 

\section{Conclusion}
In this work we demonstrated that the \ch{Zn^{2+}} doping of \ch{CsPbBr_3} \ncs{} is a a very efficient approach for enhancing the stability and brightness of LHP \ncs{}.
The studied Zn-treated \ch{CsPbBr_3} \ncs{} exhibit high single-photon purity, with \ch{g^2 (0)} values as low as $\approx 0.08$, and significantly reduced blinking behavior on a sub-millisecond time scale. Furthermore, we observed remarkable stability in both the brightness and single photon purity of the emission across a large range of excitation powers. 
This synthesis approach should also enable to achieve more stable emissions in the red and near-infrared (NIR) spectral ranges, where iodine components are prone to instability.
So far the instability of perovskite nanocrystals constituted the main drawback preventing the integration of these emitters into various nanophotonics systems, including waveguides, microcavities and metal plasmonic nanostructures. 
Our work opens up interesting prospects to couples such emitters with optimized nanophotonic architectures, enabling to Purcell enhance their emission, by increasing their radiative rate and suppressing dephasing effects. The operation of this new class of highly stable perovskite nano-emitters at cryogenic temperature is also expected to furtherly improve the recent results on the indistinguishability of the emitted photons \cite{kaplan2023hong}, a must-go for practical quantum technologies applications.

\section{Supporting information}
\begin{itemize}
    \item Methods, Material characterization, Experimental set-up, Noise cleaning of the \ch{g^2}, blinking analysis, FLID images. 
\end{itemize}

\begin{acknowledgement}

Authors thank Yoann Prado for oleylamonium bromide preparation, Angshuman Nag for his helpful insights and Chengjie Ding and Antonio Balena for the helpful discussions regarding preliminary results. This work was supported by French state funds managed by the ANR through the grant IPER-Nano2 (ANR-18CE30-0023) and by the European Union’s Horizon 2020 research and innovation program under grant agreement No 828972 - Nanobright. AB and QG are members of the Institut Universitaire de France (IUF). 
\newline
\newline
The authors declare no competing financial interest.
\end{acknowledgement}

\bibliography{achemso-demo}

\end{document}


\newpage

\section{I. Methods}

\subsection{Chemicals}
\ch{PbBr_2} (Alfa Aesar, $98.5\%$), \ch{Cs2CO3} (Alfa aesar, $99,99\%$), oleylamine (OLA, Acros, $80-90\%$), oleic acid (OA, Sigma-Aldrich), octadecene (ODE, Acros Organics, $90\%$), toluene (VWR, rectapur), Bromic acid (HBr, ABCR, $48\%$ aqueous solution), Ethyl acetate (JT baker), zinc diethyldithiocarbamate (ZnDDTC, alfa Aesar $17-19.5 \%$ in weight).

\subsection{Caesium oleate precursor}

We mix in a \SI{50}{\milli \liter} three-neck flask,  \SI{412}{mg} of \ch{Cs_2CO_3}, \SI{20}{\milli \liter} of ODE and \SI{1.25}{\milli \liter} of OA.  The content of the flask is stirred and degased under vacuum at room temperature for \SI{25}{min}. The flask is heated at \SI{110}{\celsius} for \SI{30}{min}. The atmosphere is switched to nitrogen and the temperature raised to \SI{150}{\celsius}. The reaction is carried on for \SI{30}{min}. At this stage, the salt is fully dissolved. The temperature is cooled down below \SI{100}{\celsius} and the flask degased under vacuum. Finally this solution is used as a stock solution.

\subsection{Oleylamonium bromide (\ch{OLABr})}

We mix \SI{10}{\milli \liter}  of OLA with \SI{1}{\milli \liter} of HBr in a \SI{25}{\milli \liter} three-neck flask that is then heated at \SI{80}{\celsius} under vacuum. The atmosphere is then switched to nitrogen and the temperature raised to \SI{120}{\celsius} for \SI{2}{h}. This solution is then transfered to the glove box using air free method and used as stock solution.

\subsection{\ch{CsPbBr_3} \ncs{} synthesis}
In a three neck flask, \SI{280}{mg} of \ch{PbBr2} are mixed with \SI{20}{\milli \liter} of ODE.  The flask is degased under vacuum at room temperature for \SI{15}{min}. Then, the temperature is raised to \SI{110}{\celsius}. Then \SI{2}{\milli \liter} of OLA are injected. Once the vacuum and temperature have recovered, we inject \SI{2}{\milli \liter} of OA. The solution is further degased at \SI{120}{\celsius} for \SI{30}{\min}. The atmosphere is switched to nitrogen and the temperature raised to \SI{180}{\celsius}. Around \SI{2}{\milli \liter} of CsOA solution are injected, and the solution turns yellow. The reaction is conducted for \SI{30}{s} and finally quickly cooled-down by removing the heating mantle and using a water bath. The solution is transferred to plastic tube and centrifuged. The supernatant is discarded. The pellet is dispersed in \SI{5}{\milli \liter} of toluene. The same volume of ethylacetate is added and the mixture is centrifugated again for \SI{5}{min} at \SI{6000}{rpm}. The obtained dried pellet is stored in the air free glove box.

\subsection{Zn shelling}
\SI{30}{\milli g} of \ch{CsPbBr_3} \ncs{} are mixed with \SI{20} {\micro \liter} of \ch{OLABr}, \SI{16}{mg} of ZnDDTC and \SI{3} {\milli \liter} of ODE. The mixture is then sonicated, before being transfered to a \SI{25}{\milli \liter} three-neck flask. The flask is heated at \SI{120}{\celsius}. The duration is tuned from \SI{5} to \SI{60}{min}. The color tends to swicth from green to brown green but PL remains green. Then the mixture is centrifugated. The obtained is redispersed in toluene. The same volume of ethyl acetate is added and the mixture is centrifugated again. The supernatant is discarded and the pellet dispersed in toluene.

\subsection{Absorption spectra}
For absorption, we used dilute solution of NC in hexane. The spectra were acquired using a Jasco V730 spectrometer.

\subsection{Transmission electron microscopy}
 A drop of diluted NCs solution was drop-casted onto a copper grid covered with an amorphous carbon film. The grid was degassed overnight under secondary vacuum. Imaging was conducted using a JEOL \SI{2010} transmission electron microscope operated at \SI{200} kV. Complementarily, TEM/STEM observations were made on a Titan Themis \SI{200} microscope (FEI/Thermo Fischer Scientific) equipped with a geometric aberration corrector on the probe. The microscope was also equipped with the "Super-X" systems for EDX analysis with a detection angle of \SI{0.9} steradian. The observations were made at \SI{200} kV with a probe current of about \SI{35} pA and a half-angle of convergence of \SI{17} mrad. HAADF-STEM images were acquired with a camera length of \SI{110} {\milli \meter} (inner/outer collection angles were respectively \SI{69} and \SI{200} mrad).

\subsection{X-ray photoemission measurements (XPS)}
 For photoemission spectroscopy, we used the Tempo beamline at synchrotron Soleil. Films of nanocrystals were spin-casted onto a gold coated Si substrate with an 80 nm tick gold layer. To avoid any charging effect during measurements, the ligands of the nanocrystals were exchanged by dipping the film in ethyl acetate. Samples were introduced in the preparation chamber and degassed until a vacuum below $10^{-9}$ mbar was reached. Then samples were introduced to the analysis chamber. The signal was acquired by a MBS A-1 photoelectron analyzer. Acquisition was done at constant pass energy (\SI{50} eV) within the detector. Photon energy of \SI{150} eV was used for the acquisition of valence band and work function while \SI{600} eV photon energy was used for the analysis of the core levels. A gold substrate was employed to calibrate the Fermi energy. The absolute value of the incoming photon energy was determined by measuring the first and second orders of Au4f core level peaks. Then for a given analyzer pass energy, we measured the Fermi edge and set its binding energy as zero. The same shift was applied to all spectra acquired with the same pass energy. To determine the work function, an \SI{18} V bias was applied, and its exact value was determined by observing the shift of a Fermi edge.
 
\section{II. Material characterization}

\subsection{II.1 The pristine \ch{CsPbBr_3}}

\begin{figure*}[ht!]
    \includegraphics[width=\linewidth]{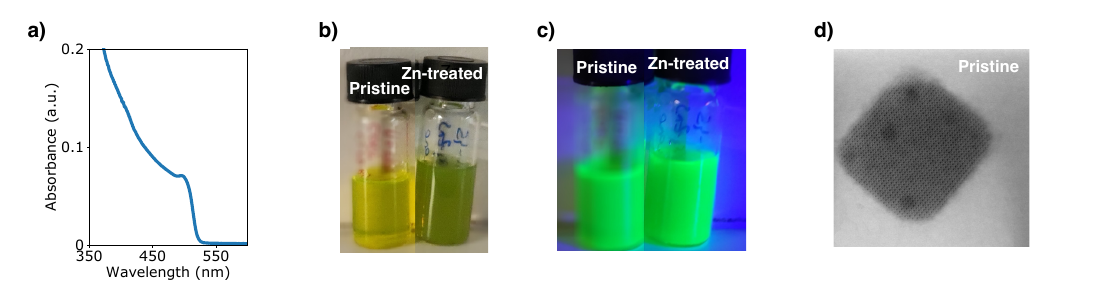}
     \caption{a) Absorption spectra for the pristine \ch{CsPbBr_3} \ncs{}. Under room light (b) and under UV illumination (c) images of the \ch{CsPbBr_3} \ncs{} before and after their exposure to Zn(DDTC) precusor. d) High resolution TEM image of a \ch{CsPbBr_3} NC. }
   \label{fig:S1}
\end{figure*}

The pristine \ch{CsPbBr_3} NCs grown using Protecescu’s procedure\cite{protesescu2015nanocrystals} depicts an absorption edge at around 520 nm (Figure\ref{fig:S1}a), giving them a yellow aspect (Figure\ref{fig:S1}b) in solution that is associated with a bright green photoluminescence (PL)(Figure\ref{fig:S1}c)). The particles present a cuboid aspect (Figure\ref{fig:S1}c). Some dark spot can be observed that have been associated with the reduction of the \ch{Pb^{2+}} by the electron beam.

In a second step, the grown cubes are exposed to the ZnDDTC precursor. During the reaction, the solution acquires a darker green aspect, but preserves the green PL. The darker aspect is likely resulting from PbS formation which is not emitting light in the spectral range of interest.

\subsection{II.2 Chemical analysis of the Zn-treated \ch{CsPbBr_3} NCs}
\begin{figure*}[ht!]
    \includegraphics[width=100mm]{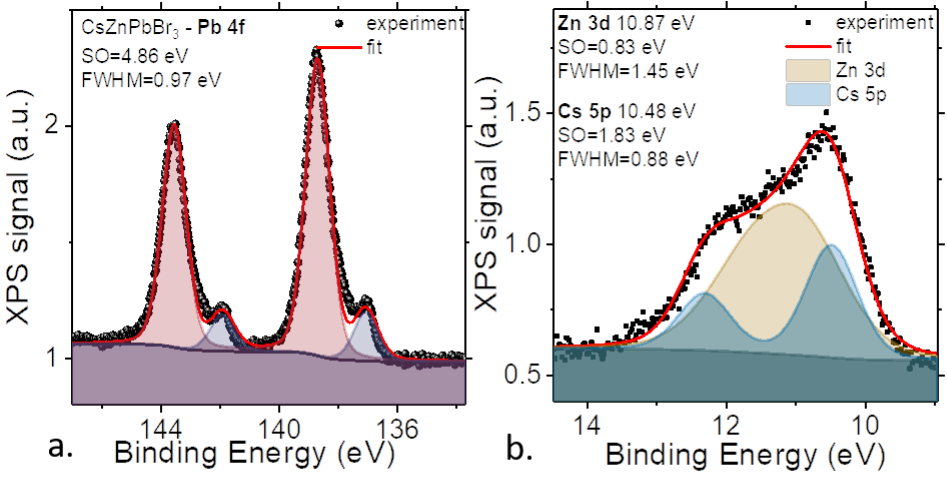}
     \caption{Core level analysis from XPS. a) Pb 4f state and b) Zn3d state for the Zn-treated
     \ch{CsPbBr3} NCs}
   \label{fig:S2}
\end{figure*}

\begin{figure*}[h!]
    \includegraphics[width=100mm]{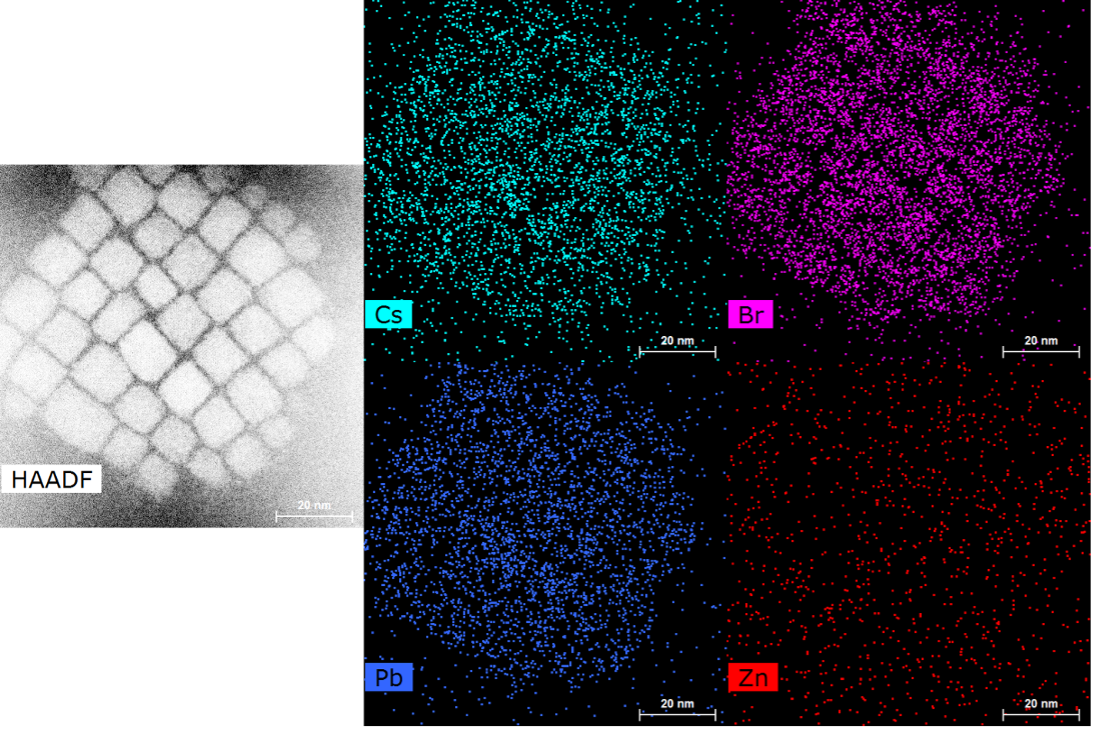}
     \caption{STEM-HAADF and EDX analysis of the Zn-treated \ch{CsPbBr_3} NCs}
   \label{fig:S3}
\end{figure*}

X-ray photoemission (XPS) is used to probe the chemical composition of the \ch{CsPbBr_3} \ncs{} after their exposure to the Zn(DDTC) precursor. The Pb 4f state displays two contributions, see Figure\ref{fig:S2}a. The main one, appearing with a 138.7 eV binding energy, is attibuted to the \ch{Pb_2^+} within the perovskite lattice. The smaller contribution appearing at lower binding energy (137 eV) is attributed to metallic \ch{Pb^0}, as already observed on the TEM image (dark spots in Figure\ref{fig:S1}b)).
The Zn presence is confirmed by XPS, see Figure S 2b. The Zn 3d state, through overlapping with Cs 5p states,present a 10.87 eV binding energy.
To further give insights on the effect of Zn exposure, we have performed energy dispersive X-ray spectroscopy in a TEM, see Figure\ref{fig:S3}. Zn content is found to be low ($0.5\%$ atomic ratio) and we see no correlation between its localization and the NC surface, suggesting that Zn mostly come as a dopant. Note that S analysis is complicated by EDX due to the overlap of the S and Pb contributions.

\newpage
\section{III. Experimental set-up}
\begin{figure*}[ht!]
    \includegraphics[width=150 mm]{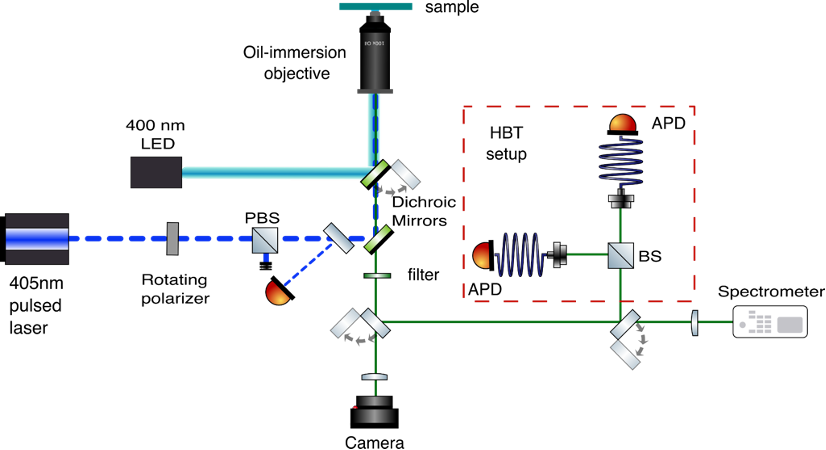}
     \caption{Experimental set-up.  Schematics of the inverted confocal microscope set-up used in the measurements, allowing both wide-field microscopy for ensamble imaging and confocal microscopy for single particle level characterization.}
   \label{fig:S4}
\end{figure*}

Figure\ref{fig:S4} illustrates the experimental setup employed for wide-field and confocal microscopy, respectively in ensemble and single particle measurements. To enable effective separation of the Zn-treated \ch{CsPbBr_3} \ncs{} and facilitate confocal microscopy, the solution containing the particles was appropriately diluted in toluene and spin-coated onto a fused-silica substrate. In the wide-field configuration, the substrate placed on the inverted microscope (Nikon Eclipse Ti) was illuminated with a 400 nm LED lamp (CooLED pE-100) and imaged using a CMOS camera (Hamamatsu ORCA-Flash 4.0) to capture videos for stability evaluation and emitter localization. Once the emitters were located, a picosecond pulsed laser (Pico Quant P-C-405B) with a 405 nm output wavelength, pulse width of less than 100 ps, and repetition rate of \SI{2.5}{\mega Hz} was used for excitation. In the confocal setup, the laser beam was focused onto the sample using a 100X oil immersion microscope objective with a numerical aperture of 1.4. The emitted photoluminescence (PL) was collected back through the same objective, spectrally filtered with a dichroic mirror and a long pass filter (cut-off wavelength of 435 nm) to eliminate the excitation laser, and then directed to various detection systems. These detection part included the CMOS camera for saturation measurements,   a spectrometer for steady-state spectroscopy, and a Hanbury Brown-Twiss setup for time-correlated single photon counting (TCSPC) detection. The TCSPC module employed in our setup (Pico Harp 300) enabled us to capture the complete fluorescence dynamics using the Time-Tagged-Time-Recorded (TTTR) method. This involved recording the arrival times of all photons relative to the beginning of the experiment (time tag), along with the picosecond TCSPC timing relative to the excitation pulses. By implementing this technique, we were able to obtain precise temporal information about the fluorescence events, allowing for detailed analysis of the second order correlation function, the photoluminesce decay and the blinking.  
All measurements were conducted at room temperature.
\newpage
\newpage
\section{IV. Noise cleaning on $g^{(2)}(\tau)$}

To accurately perform the $g^{(2)}(\tau)$ measurement, it is necessary to consider how the coincidences are counted. Let $M(\tau)$ represents the number of counts measured at a specific delay $\tau$. Each count can be generated either by a start from the or from the background signal and by a stop signal or from the background. We define $s(\tau)$ as the probability of having a start (or stop) generated by the signal and $b(\tau)$ as the probability of having the start (or stop) generated by the background. When both $b(\tau)$ and $c(\tau)$ are significantly smaller than 1, we can express this relationship as follows: 
\begin{equation}
M(\tau)=C\tonda{b(\tau)+s(\tau)}\tonda{b(\tau)+s(\tau)}
\end{equation}
where C is a constant of proportionality. We define $M'=\frac{M}{C}$, where $M'$ represents the corrected number of counts. Considering a delay $\tau_b$ between two consecutive peaks where there is no signal, we have that $s(\tau_b)=0$ and we can write:  
\begin{equation}
    M'(\tau_b)=b^2(\tau)
\end{equation}
To find $M_c(\tau)=Cs^2(\tau)$, we solve the system of equations and determine that:
\begin{equation}
M_c(\tau)=M(\tau)+M(\tau_b)-2\sqrt{M(\tau)}\sqrt{M(\tau_b)}
\end{equation}
This formula allows us to obtain the $g^{(2)}(\tau)$ histogram, which is cleaned from background counts.

\newpage 
\section{V. Blinking analysis}
\begin{figure}[ht!]
    \includegraphics[width=120 mm]{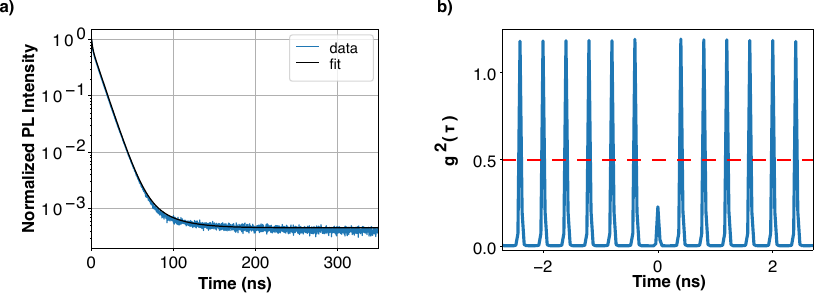}
     \caption{a) PL decay relative to the entire intensity time-trace (blue line). Data are fitted with a bi-exponential model taking into account background noise (dark line). b) Corresponding \ch{g^2($\tau$)} histogram showing antibunching.}
   \label{fig:decayg2}
\end{figure}

\begin{figure}[ht!]
    \includegraphics[width=\linewidth]{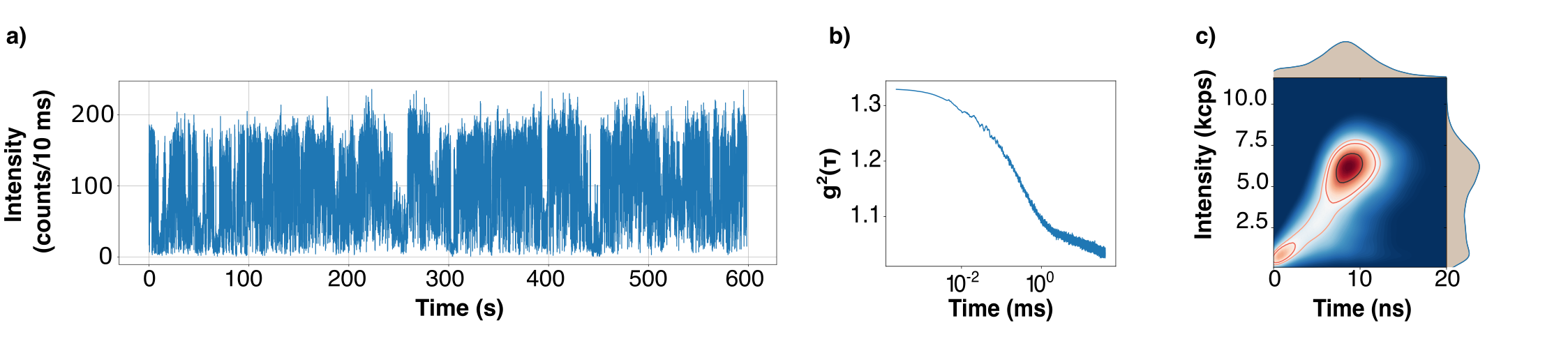}
     \caption{a) Intensity time-trace for the pristine \nc{}. b)
     \ch{g^2} function at large delays for the pristine \nc{}. c) FLID distribution for the pristine \nc{}. }
   \label{fig:g2far}
\end{figure}

In Figure\ref{fig:decayg2} we report the over-all dacay and the proof of 
antibunching relative to intensity time-trace reported in Figure4. 
In Figure\ref{fig:g2far} we show the intensity time-trace, the \ch{g^2} function at large delays and the FLID distribution in the case
of pristine \ch{CsPbBr_3} \nc{}.

\newpage
\section{VI. FLID images}

To generate the Fluorescence Lifetime Intensity Distribution (FLID) images shown in Figure5, we employed kernel density estimation\cite{rosenblatt1956remarks,parzen1962estimation}. Starting with the photoluminescence (PL) time-traces obtained with a binning time of \SI{10} {\milli s}, we calculated the main arrival time for the photons detected within each bin. This resulted in a corresponding lifetime time-trace, as depicted in Figure4c. The upper panel of Figure4c displays the PL intensity time-trace over an enlarged period of a few seconds, while the bottom panel exhibits the corresponding lifetime trajectories. The main arrival time, combined with the mean intensity of the bin, represents a point in the FLID intensity-time space.
\begin{figure*}[htp]
    \includegraphics[width=120 mm]{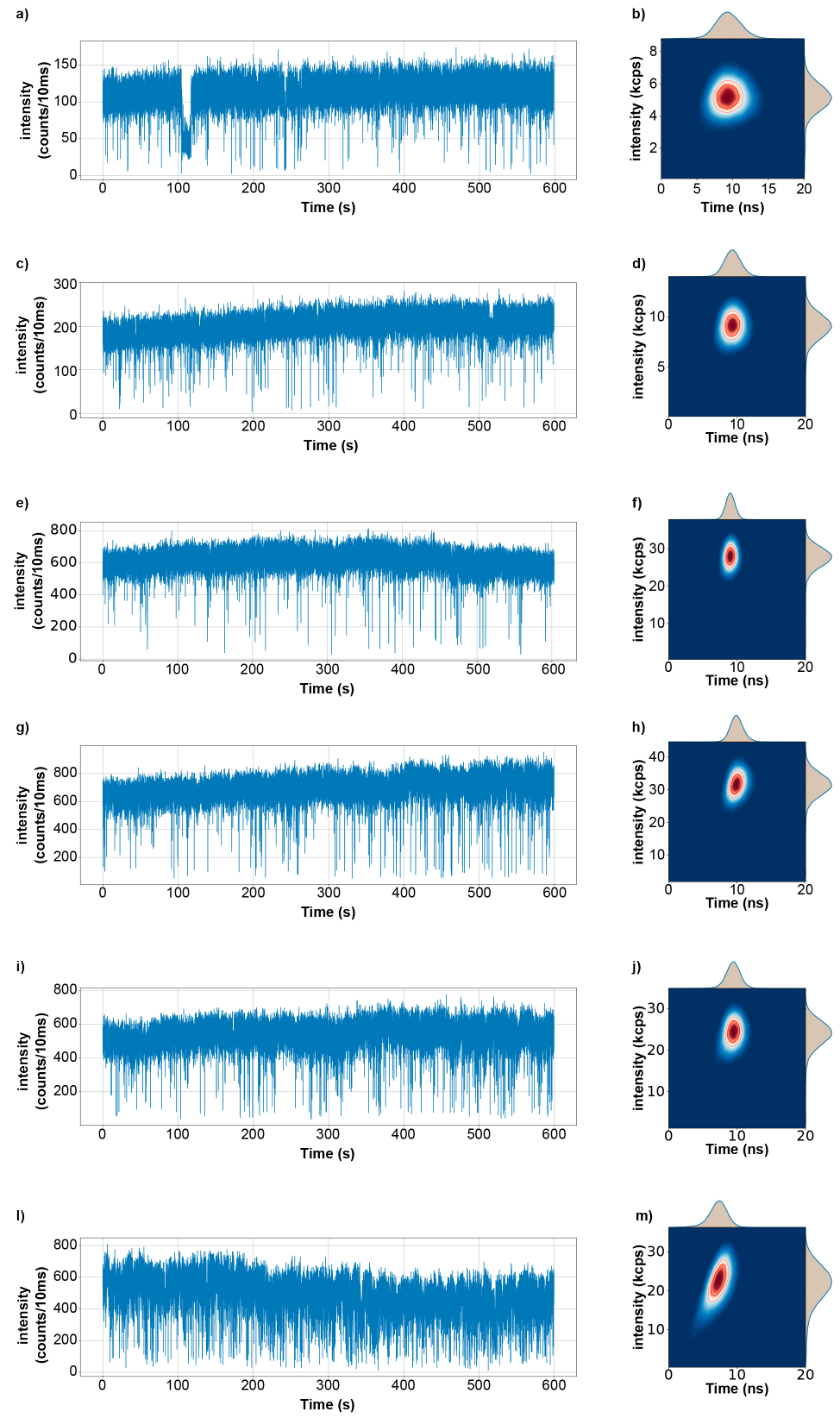}
     \caption{a-c-e-g-i-l) Intensity time-traces at different excitation powers (respectively to $0.25,0.5,1,1.5,2,3$ times $P_{sat}$) b-d-f-h-j-m) Fluorescence lifetime-intensity distribution (FLID) images of a single emitter, excited respectively at $0.25,0.5,1,1.5,2,3$ times $P_{sat}$.}
   \label{fig:powertraces}
\end{figure*}
In Figure\ref{fig:powertraces} we report the PL time-trace and the FLID images obtained at different excitation power, corresponding respectively to $0.25,0.5,1,1.5,2,3$ times $P_{sat}$.

\newpage

\bibliography{achemso-demo}